\documentclass[draft,onecolumn]{IEEEtran}

\usepackage{amsmath,amstext,amsfonts,amssymb}
\usepackage{graphicx}
\newtheorem{theorem}{\sc Theorem}

\newtheorem{lemma}{\sc Lemma}
\newtheorem{coro}{\sc Corollary}
\newtheorem{req}{\sc Requirement}
\newtheorem{nota}{\sc Notation}
\newtheorem{defin}{\sc Definition}
\newtheorem{rem}{\sc Remark}
\newtheorem{cla}{\sc Claim}
\newtheorem{ex}{\sc Example}

\newenvironment{example}{\begin{ex}}{\hspace*{\fill}$\Diamond$\end{ex}}

\newenvironment{corollary}{\begin{coro}}{\end{coro}}

\begin{document}

\bibliographystyle{plain}
\title{Tales of Huffman}
%\title{Huffman-like Coding With Multiple Length Restrictions}

% Leave blank; editors will write the exact dates above

\author{
Zvi Lotker \thanks{CWI
      {\tt email: lotker@cwi.nl}}
      \and Paul Vitanyi \thanks{CWI
      {\tt email: Paul.Vitanyi@cwi.nl}}  }
\date{}

\maketitle

\begin{abstract}
We study the new problem of
Huffman-like codes subject to individual restrictions
on the code-word lengths of a subset of the source words.
These are prefix codes with minimal expected code-word length for
a random source where additionally the code-word lengths of a subset of the
source words is prescribed, possibly differently for
every such source word. Based on a structural analysis of properties
of optimal solutions,
we construct an efficient dynamic programming algorithm for this
problem, and for an integer programming problem that may be of
independent interest.
\end{abstract}

\section{Introduction}

We are given a random variable $X$ with outcomes in a
set of $\chi =\{x_1,...,x_n\}$ of source words with associated
probabilities $P(X=x_i)=p_i$ with
$p_1\geq p_2,..., \geq p_n > 0$,
and a code word subset of
$\Omega=\{0,1\}^*$, the set of finite binary strings.
Let $l(y)$ denote the {\em length} (number of bits) in $y \in \Omega$.
Let $I=\{i: x_i \in \chi \}$.

The {\em optimal source coding problem}
is to find a 1:1 mapping $c:  \chi \rightarrow \Omega$, satisfying
$c(x_i)$ is not a proper prefix of  $c(x_j)$ for every pair $x_i,x_j \in \chi$,
such that
${\cal L}_{c}(X)= \sum_{i \in I} p_i l(c(x_i))$ is minimal among all
such mappings.
Codes satisfying the prefix condition are called {\em prefix codes}
or {\em instantaneous codes}.

This problem is solved theoretically up to 1 bit by Shannon's Noiseless
Coding Theorem \cite{Sh48}, and exactly and practically by a well-known greedy
algorithm due to Huffman~\cite{Hu52}, which for $n$ source words
runs in $O(n)$ steps,
or $O(n \log n)$ steps if the $p_i$'s are not
sorted in advance. If $c_0$ achieves the desired minimum,
then denote ${\cal L}(X) = {\cal L}_{c_0}(X)$.

We study the far more general question of length
restrictions on the individual code words, possibly different for
each code word. This problem has not been considered before. The
primary problem in this setting is the problem with equality lengths
restrictions, where we want to find the minimal expected code-word
length under the restriction of individually prescribed code-word
lengths for a subset of the code words. Apart from being a natural
question it is practically motivated by the desire to save some part
of the code tree for future code words, or restrict the lengths of
the code words for certain source words to particular values. For
example, in micro-processor design we may want to reserve code-word
lengths for future extensions of the instruction set. No polynomial
time algorithm was known for this problem. Initially, we suspected
it to be NP-hard. Here, we show an $O(n^3)$ dynamic programming
algorithm. This method 
allows us to solve 
an integer programming
problem that may be of independent interest. 
The key idea is that
among the optimal solutions, some necessarily exhibit structure that
makes the problem tractable. This enables us  to develop an
algorithm that finds those solutions among the many possible
solutions that otherwise exhibit no such structure.
Formally, we are given {\em length restrictions}
$\{l_i: i \in I\}$, where the $l_i$'s
are positive
integer values, or the {\em dummy} $\perp$,
and we require that the coding mapping $c$ satisfies $l(c(x_i)) = l_i$
for every $i \in I$ with $l_i \neq \perp$.
For
example the length restrictions $1,2, \perp , \dots , \perp$
mean that we have to set $l(c(x_1)) = 1$ and $l(c(x_2))=2$, say
$c(x_1)=1$ and $c(x_2)=01$. Then, for the remaining $x_i$'s
the coding mapping $c$ can use only
code words that start with $00$.
We assume that the length restrictions satisfy \eqref{eq.kraft} below,
the Kraft's inequality \cite{Kr49},
\begin{equation}\label{eq.kraft}
\sum_{i \in I} 2^{-l_i} \leq 1,
\end{equation}
where we take $l_i = \infty$ for $l_i = \perp$,
since otherwise
there does not exist a prefix code as required.

{\bf Related Work:}
In \cite{LH90,CD92,MTK95,BS01,KN05} a variant of
this question is studied by bounding the maximal code-word length,
which results in a certain redundancy (non-optimality)
of the resulting codes. In \cite{Ba06} both the maximal code-word
length and minimal code-word length are prescribed.

\section{Noiseless Coding under Equality Restrictions}
Shannon's Noiseless Coding Theorem \cite{Sh48} states that
if $H(X) = \sum_{i \in I} p_i \log 1/p_i$ is
the entropy of the source, then $H(X) \leq {\cal L}(X) \leq H(X)+1$.
The standard proof exhibits the Shannon-Fano code
achieving this optimum by encoding
$x_i$ by a code word $c(x_i)$ of length $l(c(x_i))= \lceil \log 1/p_i \rceil$
($i \in I$). Ignoring the upper rounding to integer values for the
moment, we see that ${\cal L}(X)=H(X)$ for a code that codes $x_i$ by a code word
of length $\log 1/p_i$. This suggests the following approach.

Suppose we are given length restrictions $\{l_i : i \in I \}$.
Let
$L = \{i \in I: l_i \neq \perp \}$
be the set of equality length restrictions,
and let ${\cal L}(X,L)$ be the minimal expected code-word length
under these restrictions given the probabilities.
Similar to Shannon's noiseless coding theorem,
we aim to  bound the minimal expected code-word
length  under equality restrictions below by an entropy equivalent
$H(X,L) \leq {\cal L}(X,L) \leq H(X,L)+1$ where $H(X,L)$
corresponds to the best possible coding with real-valued code-word lengths.
Define
\begin{equation}\label{q1}
q_i = 2^{-l_i} \textrm{ for } i \in L .
\end{equation}
If we define $q_i$'s also for the $x_i$'s with $i \in I-L$
such that $\sum_{i \in I} q_i =1$, then
altogether we obtain a new probability assignment $q_i$ for every
$x_i$ ($i \in I$), which has a corresponding Shannon-Fano code
with code lengths $l(c(x_i))= \log 1/q_i$
for the $x_i$'s.
Moreover, with respect to the probabilities induced by the original
random variable $X$, and simultaneously respecting the length restrictions,
the minimum expected code word length of such a $q$-based
Shannon-Fano code is obtained by a partition
of
\begin{equation}\label{Q1}
Q=1- \sum_{i \in L} 2^{-l_i}
\end{equation}
into $q_i$'s ($i \in I-L$) such that $ \sum_{i \in I} p_i \log 1/q_i$
is minimized. Clearly, the part $\sum_{i \in L} p_i \log 1/q_i$
cannot be improved. Thus we need to minimize
$S = \sum_{i \in I-L} p_i \log 1/q_i$ over all
partitions of $Q = \sum_{i \in I-L} q_i$ into $q_i$'s.
The partition that reaches the minimum $S$
does not change by linear scaling of the $p_i$'s. Hence we can
argue as follows.
Consider $S'= (1-Q) \log 1/(1-Q)+ \sum_{i \in I-L} q_i \log 1/q_i$
such that $S'$ is the entropy of the set
of probabilities $\{1-Q,q_i: i \in I-L\}$.
Denote
\begin{equation}\label{P1}
P=  \sum_{i \in I-L} p_i,
\end{equation}
and define
\begin{equation}\label{q2}
q^0_i = \frac{Q}{P}p_i \textrm{ for } i \in I-L.
\end{equation}
Then, both
$S$ and $S'$ with the $q_i=q^0_i$  ($i \in I-L$) 
reaches their minimum for this partition of $Q$.
\begin{lemma}\label{lem.1}
Assume the above notation with with $L,P,Q$ determined as above.
The minimal expected prefix code length under given length restrictions
is achieved by encoding $x_i$ with code length $\log 1/((Q/P)p_i)$
for all $i \in I-L$.
\end{lemma}
Let us compare the optimal expected code length under length constraints with
the unconstrained case.  The difference in code length is
\[
\sum_{i \in I} p_i \log 1/q_i - \sum_{i \in I} p_i \log 1/p_i
= \sum_{i \in I} p_i \log p_i/q_i,
\]
the Kulback-Leibler divergence $D(p \parallel q)$ between the $p$-distribution and the
$q$-distribution \cite{CT91}. The KL-divergence is always nonnegative,
and is 0 only if $p_i = q_i$ for all $i \in I$. For the optimum $q$-distribution
determined in Lemma~\ref{lem.1} for the index set $I-L$ we can compute it explicitly:
\[ \sum_{i \in I-L} p_i \log \frac{p_i}{(Q/P)p_i} = P \log \frac{P}{Q}.
\]
\begin{lemma}\label{lem.2}
Given a random source $X$ with probabilities $P(X=x_i)=p_i$ ($i \in I$),
length restrictions $\{l_i: i \in I\}$ and
with $L,P,Q$ determined as above. Then,
the minimum expected constrained code length is
\[
H(X,L) = \sum_{i \in I} p_i \log 1/p_i +
\sum_{i \in L} p_i \log \frac{p_i}{2^{-l_i}} +
 P \log \frac{P}{Q},
%\left( \sum_{i \in I-L} p_i \right)
 %\log \frac{ \sum_{i \in I-L} p_i}{1- \sum_{i \in L} 2^{-l_i}} \geq 0,
\]
which equals the minimal expected unconstrained code word length
$\sum_{i \in I} p_i \log 1/p_i$ only when $q_i=p_i$ for all $i \in I$.
\end{lemma}
Thus, the redundancy induced by the equality length restrictions is
\[
H(X,L)-H(X) = \sum_{i \in L} p_i \log \frac{p_i}{2^{-l_i}} +
 P \log \frac{P}{Q}.
\]
Note that, just like in the unconstrained case we can find a prefix
code with code word lengths $\lceil \log 1/p_i \rceil$, 
showing that the minimal expected integer prefix-code word length is in between
the entropy $H(X)$ and $H(X)+1$, the same holds for the constrained
case. There, we constructed a new set of probabilities with entropy
$H(X,L)$, and for this set of probabilities the minimal expected 
integer prefix-code word length is in between
the entropy $H(X,L)$ and $H(X,L)+1$ by the usual argument.
\begin{example}
\rm
Let us look at an example with probabilities
$X=(0.4,0.2,0.2,0.1,0.1)$ and length restrictions $L=(\perp,2,2,2, \perp)$.
The entropy $H(X) \approx 2.12$ bits, and
the, non-unique, Huffman code, without the length
restrictions, is $0,10,110,1110,1111$, which
shows the minimal integer code-word
length of ${\cal L}(X)=2.2$ bits, which is $\approx 0.08$ bits above
the noninteger lower bound $H(X)$.
The redundancy excess induced by the equality
length restrictions is $H(X,L)-H(X) \approx 0.24$ bits,
which shows that the integer minimal average code-word length ${\cal L}(X,L)$
is in between $H(X,L) \approx 2.36$ bits and $H(X,L)+1 \approx 3.36$ bits.
The actual optimal equality restricted code, given by Algorithm A below, is
$111,10,01,00,110$ with
${\cal L}(X,L) = 2.5$, which is $\approx 0.14$
bits above the noninteger lower bound
$H(X,L)$.
\end{example}

\section{Optimal Code under Equality Restrictions}
Above we have ignored the fact that real Shannon-Fano codes
have code-word length $ \lceil \log 1/p_i \rceil$ rather than
$ \log 1/p_i$. This is the reason that $H(X) \leq {\cal L}(X) \leq H(X)+1$
in the unconstrained case, leaving a slack of 1 bit for the minimal
expected code word length. The Huffman code is an on-line method to
obtain a code achieving ${\cal L}(X)$.
Gallager \cite{Ga78} has proved an upper bound on the redundancy of a
Huffman code, ${\cal L}(X)-H(X)$ of $p_n + \log [(2 \log e)/e]$
 which is approximately $p_n + 0.086$, where $p_n$ is the probability
of the least likely source message. This is slightly improved in \cite{CD92}.
Our task below is to find a Huffman-like method
to achieve the minimal expected code-word length ${\cal L}(X,L)$
in the length-constrained setting. Our goal
is to come as close to the optimum in Lemmas~\ref{lem.1}, \ref{lem.2} as is possible.

\subsection{Free Stubs}
Input is the set of source words $x_1, \ldots , x_n$
with probabilities $p_1, \ldots , p_n$  and length restrictions
$l_1 , \ldots , l_n$ that should be satisfied by the target
prefix code $c$ in the sense
that $l(c(x_i))= l_i$ for all $1 \leq i \leq n$ except for the $i$'s with
$l_i = \perp$ for which $i$'s there are no code word length restrictions.
Let $I= \{1, \ldots , n\}$,
$L= \{i : l_i \neq \perp \}$, and $P= I-L$. Denote
\[
M= |P|= |\{p_i: i \in I-L\}|
\]
For convenience in notation, assume that
the source words $x_1, \ldots , x_n$
are indexed
such that  $L= \{1,2, \ldots , k \}$
with $l_1 \leq l_2 \leq \ldots \leq l_k$.
If $l_i = l_{i+1}$ for some $i$ we set
$L:= L - \{ i,i+1 \}$;  $L := L \bigcup \{(i,i+1)\}$;
$k := k-1$;
$l_{i,i+1} := l_i -1$. We repeat this process until
there are no equal lengths
left, and finish with $l'_1 < l'_2 < \ldots < l'_{k'}$
with $l'_i$'s the resulting lengths.
That is, we just iteratively merge two
nodes which are at the same level in the tree.
Therefore, the problem is reduced to
considering a code tree with forbidden code-word
lengths $l'_i$ ($1 \leq i \leq k')$, the largest forbidden length $l'_{k'}$
leading to a forbidden node with a free sibling node at the end
of a path at the same length from the root.
That is, a code word tree with a single available free node
at each level $h$,
satisfying $1 \leq h \leq l'_{k'}$
and $h \neq l'_i$ ($1 \leq i \leq k'-1$).
Each $h$ corresponds to a path of length $h$
leading from the root to a node $n(h)$ corresponding
to a code-word prefix that is as yet unused. We call such a node (and
the path leading to it or the corresponding code-word prefix)
a {\em free stub}.
Denote the set of levels of these free stubs $n(h)$ by ${\cal H}$, and let
$m=|{\cal H}|$. Without loss of generality,
\[
{\cal H}= \{h_k : 1 \leq k \leq m\}
\]
with $h_1 < h_2 < \ldots < h_{m}$.
We now have to the find a code-word tree using only
the free stubs, such that the expected code-word length is minimized.
%If $T$ is a candidate code-word tree then $T_k$ denotes the subtree attached
%to the  free stub $n(h_k)$ at level $h_k$ with $n(h_k)$
%as its root ($1 \leq k \leq m$). 
We can do this
in the straightforward manner, dividing the probabilities in $P$ among the $m$
free stubs, and computing the minimal expected code-word length tree for the
probabilities for every stub for each of those divisions,
and determining the division giving the least expected code-word length.
We can use Huffman's construction since it doesn't depend on
the probabilities summing to 1.
There are $m^M$ possible divisions, so this process involves
computing $m^M$ Huffman trees---exponentially many unless $m=1$ which
is the unrestricted common Huffman case.

\subsection{Reduction}
Let $level(p)$ denote the number of edges in a path from the
leaf node labeled by $p$ to the root.
So $level (root)=0$. Then, if a tree is optimal
(has least expected code-word length), then
\begin{equation}\label{eq.levels}
\text{if  } level(p) < level(q) \text{ then } p \geq q,
\end{equation}
since otherwise the expected code-word length can be decreased
by interchanging $p$ and $q$. If $T$ is a prefix-code word
tree for the source words under the given length restrictions,
then the subtree $T_i$ is the subtree of $T$
with the free stub $n(h_i)$ as its root ($1\leq i \leq m$).

\begin{lemma}\label{lem.optimal}
There is a tree $T$ with minimal expected code-word length
such that if $i<j$ then $p \geq q$ for all
$p$ in $T_i$ and $q$ in $T_j$.
\end{lemma}
\begin{proof}
Suppose the contrary: for every optimal tree $T$,
there are $p<q$ with $p$ in $T_i$ and $q$ in $T_j$
for some $i<j$. Fix any such $T$.
By \eqref{eq.levels}, $level(p) \geq level(q)$.
Let $T_{p,q}$ be the subtree with root at $level(q)$
containing the leaf $p$. Then we can interchange $q$ and
$T_{p,q}$ without changing the expected code-word length represented
by the tree. This idea leads to the following sorting procedure:
Repeat until impossible:
find a least level probability, and if there are more than
one of them a largest one, that violates the condition in the
lemma, and interchange with a subtree as above.
Since no probability
changes level the expected code-word length stays invariant.
In each operation a least level violating probability moves
to a lower indexed subtree, and the subtree it is interchanged with
does not introduce new violating probabilities
at that level. 
The transformation is
iteratively made from the top to the bottom of the tree.
This process must terminate with an overall
tree satisfying the lemma,  since there
are only a given number of probabilities and indexed subtrees.
\end{proof}
For ease of notation we now assume that the source words $x_1 , \ldots , x_n$
are indexed such that the unrestricted source words are indexed
$x_1 , \ldots , x_M$ with probabilities
$p_1 \geq p_2 \geq
\ldots \geq p_M $.
The fact that we just have to look for a partition
of the ordered list of probabilities into $m$ segments,
rather than considering every choice of $m$ subsets of the
set of probabilities,
considerably reduces the running time to find an optimal prefix code.
Consider the ordered list $p_1 \geq \cdots \geq p_M$. Partition it
into $m$ contiguous segments, possibly empty, which gives
${{M+m-1} \choose {m-1}}$ partitions.
We can reduce this number by noting that in an optimal tree
the free stubs are at different heights, and therefore each of them
must have a tree of at least two elements until they have empty trees
from some level down. Otherwise the tree is not optimal since
it can be improved by rearranging the probabilities.
Therefore, we can restrict attention to partitions
into $\leq m$ segements, which contain at least two elements.
There are at most
${{M-m} \choose {m}}$ such partitions.
For each choice, for each set of probabilities
corresponding to a $i$th segment construct the Huffman tree
and attach it to the $i$th free stub, and compute the
expected code-word length for that choice. A tree associated
with the least expected codeword length is an optimal tree.
Thus, we have to construct at most ${{M-m} \choose {m}}$ Huffman trees,
which is polynomial in $M$ for fixed $m$, and also polynomial in $M$
for either $m$ or $M-m$ bounded by a constant.

\subsection{Polynomial Solution}
From each partition of $p_1 \geq \ldots \geq p_M$ into $ m$
segments (possibly empty) consisting of probabilities $P_k$ for the $k$th
segment, we can construct trees   $T_1, T_2, \ldots , T_m$
with $T_k$ having the probabilities of $P_k$ as the leaves, and
free stub $n(h_k)$ as the root. Clearly, if $T$ is an overall tree
with minimum expected code-word length, then
each subtree with the free stub $n(h_k)$ as root
considered in isolation,
achieves minimal expected code-word length over the probabilities
involved.  We want to find the optimal partition
with a minimum amount of work.
Note that, from some $s \leq m$ on, every subtree
$T_k$ with $s < k \leq m$ may be empty.

For every tree $T$, not necessarily optimal, let $L_T$ denote the expected
code-word length for the probabilities in $P$ according to tree $T$.
Define $H[i,j,k]$ to be the {\em minimal}
 expected code-word length of the leaves
of a  tree  $T[i,j,k]$
 constructed from probabilities $p_i, \ldots , p_j$ ($i \leq j$)
and with a  singlefold path from the root of $T$ to
the free stub node $n(h_k)$, and subsequently branching out to encode
the source words (probabilities) concerned. Then,
\begin{equation}\label{eq.hij}
H[i,j,k] = \sum_{i \leq r \leq j} p_r (h_k+l(p_r))
\end{equation}
each probability $p_r$ labeling a leaf at the end of
a path of length $h_k +l(p_r)$
from the root of $T$, the first part of length $h_k$
to the free stub $n(h_k)$,
and the second part of length $l(p_r)$
from $n(h_k)$ to the leaf concerned.
For a partition of the probability index sequence
$1, \ldots , M$ into $m$ (possibly empty)
contiguous segments $[i_k,j_k]$ ($1 \leq k \leq m$),
inducing subtrees $T[i_k,j_k,k]$ using free stubs $h_k$
accounting for expected code-word length $H[i_k,j_k,k]$, we obtain
a total expected code-word length for the overall tree $T$ of
\[
L_T = \sum_{1 \leq k \leq m}
H[i_k,j_k,k].
\]
Let us now  consider the expected code word length of a tree $T'$ which
consists of tree $T$ with a subset of subtrees $T_k$ removed
and the corresponding probabilities from the overall
probability set $P$. Removing subtree $T_k$ is equivalent to removing
the corresponding free stub $n(h_k)$, and turning it into a length
restriction.
\begin{lemma}
Let $T$, $P$ and $T'$ be defined as above. Let $T$ has minimal total code-word
length for $P$ then the total code-word length of every $T'$ as
above cannot be improved by another partition of the probabilities
involved among its subtrees.
\end{lemma}
\begin{proof}
({\em If}) If we could improve the total code word length of $T$ by a
redistribution of probabilities among the subtrees attached to
the free stubs then some $T'$ would not have minimal total code-word
length before this redistribution. 

({\em Only if}) If we could improve the total code-word
length of any $T'$ by redistribution of the probabilities
among its subtrees attached to the free stubs involved,
then we could also do this in the overall tree $T$ and
improve its overall total code-word length, contradicting minimality.
\end{proof}

\begin{corollary}\label{cor.seqopt1}
If tree $T$ has minimal total code-word length, then every tree $T'$
obtained from it as above has minimal total code word length.
\end{corollary}
%
%This property means in particular that a $T'$ containing $T_1, \ldots , T_k$
%and a tree $T''$ containing $T_{k+1}$ both
%have minimal code-word length iff a $T'''$ containing $T_1, \ldots , T_{k+1}$
%has minimal total code-word length. That is,
%
%\begin{corollary}\label{cor.seqopt2}
%Partition an initial segment of the probabilities $p_1, \ldots, p_M$
%into subsets of probabilities $P_1, \ldots , P_m$, possibly empty
%from some index onwards, and let
%$T'$ be the tree obtained by attaching subtree $T_k$ with
%the probabilities in $P_k$ as leaves to free stub $n(h_k)$ ($1 \leq k \leq m$).
%If $T_1 , \ldots , T_{k+1}$ rooted at these levels
%give minimal expected code-word length for
%probabilities $P_1 \bigcup \cdots \bigcup P_{k+1}$ then this minimum
%equals the minimum expected code-word length of (possibly different)
%$S'_1 , \ldots , S'_{k}$ for
%probabilities $P_1 \bigcup \cdots \bigcup P_{k}$ plus the minimum expected
%code-word length of $S'_{k+1}$ based on probabilities $P_{k+1}$,
%rooted the same way at $n(h_1), \ldots , n(h_{k+1})$, respectively.
%\end{corollary}
This suggests a way to construct an optimal $T$ by examining every
$k$-partition corresponding to a candidate set of $k$ subtrees (for
$k:= 1, \ldots , m$), of every initial segment $p_1 \geq \ldots \geq
p_j$ of the probability sequence (for $j:=1, \ldots , M$). The
minimal expected code-word length tree for the $k$th partition
element is attached to the $k$th free stub. The crucial observation
is that by Corollary~\ref{cor.seqopt1} the minimal total code word
length for probabilities  $P_1 \bigcup \cdots \bigcup P_{k+1}$ using
free stub  levels $h_1, \ldots , h_{k+1}$ is reached for a binary
split in the ordered probabilities and free stubs involved,
consisting of the minimal total code-word length solution for
probabilities $P_1 \bigcup \cdots \bigcup P_{k}$ using stub levels
$h_1, \ldots , h_{k}$ and probabilities $P_{k+1}$ using free stub
level $h_{k+1}$. Computing the optimal minimum code-word lengths of
initial probability segments and initial free stub level segments in
increasing order, this way we find each successive optimum by using
previously computed optima. This type of computation of a global
optimum is called dynamic programming. The following Algorithm A
gives the precise computation. At termination of the algorithm, the
array $F[j,k]$ will contain the minimal expected code word length of
a tree $T[1,j,k]$ using the  largest $j$ ($j \leq M$) probabilities
$p_1, \ldots , p_j$, optimally divided into subtrees attached to the
least level $k$ ($k \leq m$) free stubs. Thus, $T[1,j,k]$ contains
subtrees $T[j_i,j_{i+1},i+1]$ ($0 \leq i < k$, $j_0 = 1 \leq j_i
\leq j_{i+1} \leq j_k$) each such subtree with $n(h_{i+1})$ as the
root ($0 \leq i < k$). Thus, on termination $F[M,m]$ contains the
minimal expected code-word length of the desired optimal tree $T$
with subtrees $T^0_1, \ldots , T^0_m$ with $T^0_k$ rooted at the root of
$T$ and using free stubs $n(h_1), \ldots , n(h_m)$, respectively.
(Note, $T^0_k$ consists of the previously defined subtree $T_k$,
rooted at free stub $n(h_k)$, plus the path from $n(h_k)$ to the
root of $T$.) We can reconstruct these subtrees, and hence the
desired code words, from the values of the array $s$ on termination.
Denote $s_{m+1}=M$, $s_m = s(M,m)$, $s_k = s(s_{k+1},k)$ ($m \geq k
\geq 2$), and $s_1 = 0$. Then, $T^0_k = T[s_k+1,s_{k+1},k]$ and
$F[s_{k+1},k]$ equals the expected code-word length of the source
words encoded in subtrees $T^0_1, \ldots , T^0_k$ ($1 \leq k \leq m$).
Thus, the array values of $s$ give the desired partition of the
ordered probability sequence $p_1 \geq \cdots \geq p_M$, and we can
trivially construct the tree $T$ and the code-words achieving
minimal expected code-word length by Huffman's construction on the
subtrees.
\subsection*{Algorithm A}
{\bf Input:} Given $n$ source words with ordered probabilities and
equality length restrictions, first check whether \eqref{eq.kraft}
is satisfied with $\perp = \infty$, otherwise return ``impossible''
and quit. Compute
free stub  levels $h_1 < \cdots < h_m$.
probabilities $p_1 \geq \cdots \geq p_M$ as above.
\begin{tabbing}
\= {\bf Step 1:} \=  Compute \= $H[i,j,k]$ \=  as in  \eqref{eq.hij},\= 
 for all  $i,j$ and $k$ %\\
%\>  \> \>  
($1 \leq i \leq j \leq M$, $1 \leq k \leq m$).\\

\> {\bf Step 2:}  Set $F[j,1] := H[1,j,1]$
($1 \leq j \leq M$).\\

\> {\bf Step 3:}  {\bf for }   $k := 2, \ldots , m$ {\bf do} \\

    \>  \>  \>{\bf for }   $j := 1 , \ldots , M$ {\bf do}\\

    \>  \>  \>$F[j,k] := \min_{i: 1 \leq i \leq j}  \{ F[i, k-1] $ %\\
      %\>  \>  \>  \>  \> 
 $+ H[i+1,j,k] \}$; \\
    \>  \>  \>$s[j,k] := i_0$, with $i_0$ the least $i$ %\\ 
      %\>  \>  \>  \>  
achieving the minimum\\

\end{tabbing}
{\bf end of Algorithm}
\begin{theorem}
Given $n$ source words with probabilities and equality length restrictions,
Algorithm A 
 constructs a prefix code with optimal expected code word length
in $O(n^3)$ steps.
\end{theorem}
\begin{proof}
The correctness of the algorithm follows from
Corollary~\ref{cor.seqopt1} and the discussion following it.

The complexity of computing the $h_1 < \cdots < h_m$
and $p_1 \geq \ldots \geq p_M$ is $O( n \log n)$.
Step 1 of the algorithm takes
$O(M^3)+O(m)$ steps. First, compute for every $i,j$
the quantities $P[i,j] := \sum_{i=r}^j l(p_r)$,
$L[i,j] := \sum_{i=r}^j p_r l(_r)$. There
are $O(M^2)$ such quantities and each computation takes $O(M)$ steps.
Second, for every $k$ compute $H[i,j,k] = L[i,j]+h_k P[i,j]$.
There are $m$ such quantities and each computation takes $O(1)$ steps.
Step 2 of the Algorithm takes $O(M)$ steps.
Step 3 of the Algorithm involves a outer loop of length $m-1$,
an inner loop of length $M$, and inside the nesting the determining
of the minimum of $\leq M$
possibilities; overall $O(mM^2)$ steps. The running time of the
algorithm is therefore $O(n \log n)+O(M^3) + O(mM^2)$ steps.
Since $M,m \leq n$ this shows the stated running time.
\end{proof}

\end{document}